# Atomic-scale Deformation Process of Glasses Unveiled by Stress-induced Structural Anisotropy

Jie Dong[1,3#], Hailong Peng[2#], Hui Wang[4#], Yang Tong[4,5]*, Yutian Wang[1], Wojciech Dmowski[4], Baoan Sun[1,3]*, Takeshi Egami[4,6,7], Weihua Wang[1,3], Haiyang Bai[1,3]*

[1]*Institute of Physics, Chinese Academy of Sciences, Beijing 100190, China*

[2]*School of Materials Science and Engineering, Central South University, Changsha, Hunan 410083, China*

[3]*Songshan Lake Materials Laboratory, Dongguan, Guangdong 523808, China*

[4]*Department of Materials Science and Engineering, University of Tennessee, Knoxville, TN 37996, USA*

[5]*Institute for Advanced Studies in Precision Materials, Yantai University, Yantai, Shandong 264005, China*

[6]*Department of Physics and Astronomy, University of Tennessee, Knoxville, TN 37996, USA*

[7]*Materials Science and Technology Division, Oak Ridge National Laboratory, Oak Ridge, TN 37831, USA*

#These authors contributed to the paper equally

*Correspondence to: ytong1@utk.edu; sunba@iphy.ac.cn; hybai@iphy.ac.cn

**Abstract**

Experimentally resolving atomic-scale structural changes of a deformed glass remains challenging owing to the disordered nature of glass structure. Here, we show that the structural anisotropy emerges as a general hallmark for different types of glasses (metallic glasses, oxide glass, amorphous selenium, and polymer glass) after thermo-mechanical deformation, and it is highly correlates with local nonaffine atomic displacements detected by the high-energy X-ray diffraction technique. By analyzing the anisotropic pair density function, we unveil the atomic-level mechanism responsible for the plastic flow, which notably differs between metallic glasses and covalent glasses. The structural rearrangements in metallic glasses are mediated through cutting and formation of atomic bonds, which occurs in some localized inelastic regions embedded in elastic matrix, whereas that of covalent glasses is mediated through the rotation of atomic bonds or chains without bond length change, which occurs in a less localized manner.

Glasses cover various kinds of vitrified solids including oxide glasses, polymer glasses, and metallic glasses (MG), which have been an indispensable part of our daily life and modern world[1]. Despite their different nature of atomic bonding and internal dynamics, glassy materials exhibit remarkably similar deformation behaviors[2,3]. For example, most glasses display high strength, yet tend to show strain localization upon shear[4], and hence exhibit brittle behavior[5,6]. These deformation features are fundamentally related to the unique atomic-scale deformation mechanism of glasses for which the lattice defects do not exist in the amorphous structure. Over the past decades, numerous atomic simulations showed that the plastic flow in glasses at the atomic scale involves local structural rearrangement at defect-like regions[7,8]. Many theories were proposed based on this atomistic picture, including the free volume model[9], shear transformation zones model[10-12], *etc*, which have been partially verified by some simulated experiments such as bubble raft experiments[13] and imaging particle motions in shearing colloid glasses[14]. Nevertheless, experimentally resolving atomic-scale structural arrangements in atomic or molecular glasses remains a significant challenge. Furthermore, despite the success on local arrangement modeling, identification of specific atomic mechanisms for different types of glasses is still not fully achieved. For example, covalently bonded glasses exhibit plastic flow behavior that significantly differs from that of metallic glasses[15,16]; high-density components are strikingly found to act as "plasticity carriers" responsible for the plastic flow in the network-bonded silicon glass[15], which is opposite of the picture for metallic glasses. However, we lack the fundamental understanding of the atomistic origin of such contrasting behavior, in terms of the atomic-level process responsible for the plastic flow in various kinds of glasses.

We propose the structural anisotropy induced by stress as an indicator for such a generalized signature. Glasses should be structurally isotropic at an ideal state. Conventional structural analysis of glasses such as X-ray diffraction based on the structural isotropy, are not sensitive to the atomic-scale structural rearrangement induced by deformation. Structural anisotropy could emerge in some cases, such as the melt-spun ribbons under directional shearing flow[17] or the deposited glassy films by directional growth[18]. The observed anisotropy is also suggested as the structural response to the mechanical deformation or stress, which has been reported in metallic glasses [19-21]. If deformation is elastic and uniform, the stress-induced anisotropy can be well theoretically modeled and the isotropy-to-anisotropy transition is reversible[19,20]. Yet, in practice, the stress-induced anisotropy still persists even after the removal of elastic stress, reflecting the intrinsic structural heterogeneity and viscoelastic nature of glasses. Although the atomic origin of stress-induced anisotropy remains elusive, it is definitely related to the rearrangement of atoms during deformation and can provide important clues for the atomic-scale deformation process of glasses.

We chose 4 different types of glasses for the structural anisotropy study, including one metallic glass ($Zr_{52.5}Cu_{17.9}Ni_{14.6}Al_{10}Ti_5$, Vit105), a polymer glass (polystyrene, PS), a monatomic glass (amorphous Se) and an oxide glass ($B_2O_3$). These glasses were first annealed at the temperature of $0.9T_g$ ($T_g$ is the glass transition temperature), and then subjected to creep treatments at $0.9T_g$ and a constant uniaxial compressive stress of

$0.9s_y$ ($s_y$ is the yield strength of the glass at $0.9T_g$). For more experimental details, one can see **Text SI** in **Supplementary Materials (SM)**[22]. After the creep treatment, independent high-energy X-ray diffraction measurements were carried out for each sample in two orientations, as illustrated in **Fig. 1a**. The first measurement was performed with the instrumental $z$-axis parallel to the loading axis of the crept samples (denoted as P orientation). Then, the sample was in-plane rotated by 90° around the diffraction axis (denoted as N orientation), and the measurement was repeated. Two-dimensional (2-D) diffraction patterns of samples were collected for each orientation. The existence of structural anisotropy in the crept samples can be justified by different 2-D diffraction patterns for P and N orientations. This procedure was adopted to remove accidental anisotropy in the detector and the experimental setup. For glassy materials with isotropic structure, the circular diffraction rings obtained from two states cancel each other after the difference, whereas the diffraction rings for anisotropic glassy solids are distorted, which do not cancel after rotating 90°. Typical examples of differential diffraction patterns for both the annealed and crept Zr-based MGs are shown in **Fig. 1b** and **c**, respectively. One can see that the annealed MG has an ideal isotropic atomic structure with a homogeneous contrast on the differential diffraction pattern, while the crept MG shows non-overlapped ellipse diffraction rings, indicating the existence of anisotropic atomic structure. The same phenomenon is also observed for other types of glasses (see **Fig. S1** in **SM**). These results suggest that structural anisotropy is a general signature of glasses of various types as driven by the plastic deformation during the creep.

Next, we perform a quantitative structural analysis on the anisotropic diffraction patterns. For anisotropic glassy materials, spherical harmonic expansion is employed to separate the isotropic part of the structure function, $S_0^0(Q)$, and the anisotropic part,

$S_2^0(Q)$, from the measured total structure function, $S(Q)$[19,23]

$$S(\mathbf{Q}) = \sum_{l,m} S_l^m(Q) Y_l^m\left(\frac{\mathbf{Q}}{Q}\right). \tag{1}$$

Here, $Y_l^m$ is the spherical harmonics, $\mathbf{Q}$ is the diffraction vector with $Q = 4\pi \sin\theta/\lambda$, $\theta$ is the diffraction angle and $\lambda$ is the x-ray wavelength. Then, they are converted to the isotropic pair distribution function (PDF), $g_0^0(r)$, and the anisotropic PDF, $g_2^0(r)$, through the spherical Bessel transformation, to examine the local structural change in real space[24]:

$$g_l^m(r) = \frac{i^l}{2\pi^2 \rho_0} \int S_l^m(Q) J_l(Qr) Q^2 dQ, \tag{2}$$

where $\rho_0$ is the number density of atoms, and $J_l(x)$ is the spherical Bessel function. Notice that the isotropic PDF, $g_0{}^0(r)$ reflects radial structural changes, and the anisotropic PDF, $g_2{}^0(r)$, represents the pure shear structural changes. For a solid upon affine deformation, it has been proved theoretically that the anisotropic PDF is proportional to the first derivative of the isotropic PDF[19,20]. For uniaxial compression[25], the $g_2^0(r)$ can be expressed in terms of $dg_0^0(r)/dr$:

$$g_{2,aff}^0(r) = \varepsilon_{aff} \left(\frac{1}{5}\right)^{1/2} \frac{2(1+v)}{3} r \frac{dg_0^0(r)}{dr} \tag{3}$$

where $v$ is the Poisson's ratio and $\varepsilon_{aff}$ is the affine strain.

The left panel of **Fig. 2** compares the anisotropic PDF $g_{2,exp}^0(r)$ (the solid line) obtained from experimental results for different types of glasses, together with the calculated anisotropic $g_{2,aff}^0(r)$ (the dashed line) according to affine deformation. The comparison of isotropic $g_{0,exp}^0(r)$ and anisotropic $g_{2,exp}^0(r)$ are given in the right panel of **Fig. 2**. For Vit105 MG (**Fig. 2a**), one can see that $g^0_{2,exp}(r)$ and $g^0_{2,aff}(r)$ are matched well with each other at large interatomic distance $r > 11$ Å, indicating an affine deformation at long distance. At $r < 11$ Å, $g^0_{2,exp}(r)$ and $g^0_{2,aff}(r)$ are in phase, but $g^0_{2,exp}(r)$ has the lower peak height than $g^0_{2,aff}(r)$, indicating the occurrence of nonaffine strain or inelastic strain for $r < 11$ Å. These results are similar to the earlier reports[30,31,34], and imply that the MG deforms in a manner of localized inelastic regions embedded in elastic deformed matrix. The critical distance $r_c$ at which $g^0_{2,exp}(r)$ begin to deviate from $g^0_{2,aff}(r)$ can be roughly taken as the average size of local regions (the diameter if the region is spherical) undergoing inelastic strain, or in other words, STZs. The average diameter of STZs determined for two MGs by this way is around 2.2 nm, consistent with the values reported in literature[26,27]. From **Fig. 2b**, one can see a clear peak position shift in $g^0_{2,exp}(r)$ as compared to $g^0_{0,exp}(r)$.

It is interesting to test that whether the microscopic deformation mechanism observed in MGs can be extended to the other types of glasses or not. For nonmetallic glasses ($B_2O_3$, Se and PS), the anisotropic profiles exhibit significant difference from those in MGs. For Se $g^0_{2,exp}(r)$ and $g^0_{2,aff}(r)$ show little resemblance (**Fig. 2c**). For $B_2O_3$ and PS $g^0_{0,exp}(r)$ and $g^0_{2,aff}(r)$ become almost flat beyond 5 Å, whereas $g^0_{2,exp}(r)$ shows some oscillations above 5 Å. Their peaks of $g^0_{2,exp}(r)$ and $g^0_{2,aff}(r)$ are clearly not in phase with each other (**Fig. 2e** and **g**). This absence of similarity between $g^0_{2,exp}(r)$ and $g^0_{2,aff}(r)$ implies that the structural changes in these glasses are intrinsically non-affine. Even though they appear elastic at a macroscopic scale, at the atomic scale they undergo non-affine deformation upon anelastic deformation. This point makes them fundamentally different from MGs with respect to the mechanism of anelastic deformation.

On the other hand, the comparison of $g_{2,exp}^0(r)$ with $g_{0,exp}^0(r)$ shows their peaks are in phase (see the arrows in **Fig. 2d**, **f**, and **g** for $B_2O_3$, Se, and PS glasses, respectively). This means that the bond lengths between atoms and inter-chain distances in these glasses do not change during local non-affine atomic arrangements, suggesting that the structural changes occur due to local rotation of bond direction, without changing the bond length.

To verify the experimental results, we performed molecular dynamics (MD) simulations on the creep of a typical MG ($Cu_{50}Zr_{50}$), and a covalently-bonded polymer glass (polystyrene, PS). The creep curves and the comparison of $g_2^0(r)$ with $g_0^0(r)$ and $dg_0^0(r)/dr$ in simulation agree well with that in experiments (see **Fig. S2** in the **SM**). We calculated $g_2^0(r)$ for each particle and compared it with the non-affine displacement, $D^2$, for the same particle. As shown in **Fig. 3a** and **b**, the magnitude of the anisotropy increases with $D^2$ at any distance $r$. The correlation is more evident at the first and second nearest neighbor shells (see the arrows labeled with $r^1_{max}$ and $r^2_{max}$ in the figures). At large distances, anisotropy is prominent only at larger $D^2$, suggesting the spatial range of structural anisotropy depends linearly on the strength of non-affine deformation. We, thus, conclude that the structural anisotropy is a good measure of the magnitude and spatial extension of local plasticity both in atomic and molecular glasses.

In order to elucidate the atomic-level deformation mechanism alluded by the experimental results, we analyzed the bond breaking after the creep in metallic glasses. The bond-breaking percentage for each particle, defined as the broken bonds over the total number of the nearest neighbors, is plotted against $D^2$, averaged over the particle for each value of $D^2$, is shown in **Fig. 3c**. The number of breaking bonds monotonically increase with $D^2$, exhibiting a linear relation at small non-affine deformation. Interestingly, a linear fit hits the y-axis at a non-zero value, indicating a threshold of bond-breaking percentage of about 10% for plastic event. This means that at least one bond has to be broken around each particle for deformation to occur, because the number of nearest neighbors, the coordination number, is around 12 – 14 for metallic glasses. Thus this is a very reasonable result.

In polymer glasses, the relevant atomic motion is the rotation of covalent bonds. There are two kinds of carbon-carbon bonds in polystyrene glasses: one is C-C bonds in the aliphatic moiety characterized by its bond orientation, and the other one is the bonds in benzene rings characterized by the normal direction of the benzene rings. We calculated the distribution-probability difference after the creep for these two orientations relative to the compression axis. As shown in **Fig. 3d**, a significantly enhanced probability can be seen at $\varphi = 0°$ for the benzene rings, while the enhanced probability locates at $\varphi = 90°$ for the C-C bonds in the aliphatic moiety. This gives the evidence that compression reorients benzene rings toward the plane perpendicular to the stress, which is also a reasonable result.

A picture of atomic-level processes associated with plastic deformation in metallic and polymer glass is shown in **Fig. 4**. Plastic deformation represented by the non-affine displacement field is strongly heterogeneous in space both in metallic and polymer glasses, as shown in **Fig. 4a** and **c**. An intuitive deformation mechanism is given in **Fig. 4b** and **d** that presents configuration changes of the deformation units of the simulated glasses. Upon compression, local atomic structure in MG is plastically deformed with bond shrinking and stretching (**Fig. 4b**), resulting in bond breaking and reformation denoted by arrows, resulting in losing and gaining of neighbor atoms. On the other hand polymer glasses (**Fig. 4d**) show a quite different mechanism: the molecular chain rotates, as denoted by arrows, without obvious changes in bond length. The structural anisotropy of samples after creep measured by diffraction clearly suggests these deformation mechanisms.

Currently, the general understanding of the deformation mechanism of glasses is achieved mainly through theoretical models and simulations[9-11]. Experimentally resolving the atomic-scale deformation process of glasses with conventional techniques, such as electron microscopy, remains extremely challenging. Our results suggest that the structural anisotropy induced by creep can act as a general signature for the plastic flow of various types of glasses: They reveal hidden information about their atomic-scale deformation process, and provide a "fingerprint" of deformation mechanism. The commonly proposed deformation "order parameter" is usually based on the pre-knowledge of local packing geometry[11,28,29]. Our approach does not require any prior detailed knowledge of atomic structure, and provide the direct measure of deformation. Our simulations support the intuitive interpretation of the results, and demonstrate that the local structural anisotropy correlates with the nonaffine displacement of atoms. This approach allows direct access to the atomic-scale deformation information by diffraction experiments of samples undergone creep deformation, and hence serves as an effective tool to sort out different modes of atomic-scale plastic deformation process in glassy solids

Based on structural-anisotropy analysis, we can see that the plastic flow of MGs proceeds in a manner of localized inelastic regions embedded in elastically deformed matrix. This picture is consistent with local atomic arranging models proposed[10],[30]. The average size of local inelastic regions determined for two MGs from our analysis, is ~1 nm, consistent with the estimate by strain-rate jump experiments[26,27] and with earlier results[41]. The size of the local deformation unit is small, involving 5 atoms in average[42]. Thus, creep deformation must involve cascade process associated with ductile deformation[43]. On the other hand, surprisingly long-range elastic field is not visible for covalent and polymer glasses. These glasses are not close-packed but have more open structures. Thus, deformation occurs through local rotation of chains and covalent units. Because of the complexity of the structure the isotropic PDF, $g^0_{0,exp}(r)$, quickly loses signal with distance, making the long-range stress field invisible to the PDF. It is possible that chemically resolved PDF offers more information. It is interesting to note that $g^0_{2,exp}(r)$ shows more oscillations than $g^0_{0,exp}(r)$, and modeling this will help understand the deformation mechanism better. Rotation of atomic bonds or molecular chains is consistent with anomalous plastic flow behavior observed in directionally bonded glasses[15,16].

This research was supported by the National Key Research and Development Plan (Grant No. 2018YFA0703603), the National Natural Science Foundation of China (NSFC) (Grant No. 52192601, 52192602, 11790291, 61888102, 52001272), the Strategic Priority Research Program of Chinese Academy of Sciences with Grant No. XDB30000000. WD and TE were supported by the US Department of Energy, Office of Science, Basic Energy Sciences, Materials Sciences and Engineering Division.

**Figure Captions**

**Fig. 1.** (a) The illustration of detecting the anisotropic structure for glasses with high-energy X-ray diffraction. (b) The circle diffraction rings of isotropic structure are balanced with each other after difference for a typical as-cast glass (Vit105) sample. (c) The nonoverlapped elliptical diffraction rings induced from the anisotropic structure of the sample crept at high temperature.

**Fig. 2.** The comparison of the observed anisotropic PDF $g^0_{2,exp}(r)$ to the calculated anisotropic PDF $-\alpha * r * dg^0_0(r)/dr$, and to the observed isotropic PDF $g^0_{0,exp}(r)$, for different glasses after creep deformation: (a) and (b) MG(Vit105); (c) and (d) monoplasmatic glass (Se); (e) and (f) Polymer glass (Polystyrene); (g) and (h) Oxide glass ($B_2O_3$).

**Fig. 3.** The variation of the intensity of $g^0_{2,exp}(r)$ with the local non-affine displacment $D^2$ and the atomic distance $r$ for MG (a) and polymer glass (b) in simulations. (c) The number of broken bonds versus the $D^2$ in the simulated MG. (d) The distribution-probability difference $\Delta P(\varphi)$ for the orientation relative to the loading axis for the C-C bonds and Benzene rings, respectively, in the simulated polymer glass.

**Fig. 4.** (a) The non-affine atomic displacements (represented by the color bar) in the simulated $Cu_{50}Zr_{50}$ MG; (b) A deformed cluster in the simulated MG clearly showing changes of bond lengths, as well as a loss and gain of atoms via bonds breaking and formation. (c) The non-affine atomic displacement in the simulated polymer glass is relatively uniform compared to that in MG. (d) a chain of the simulated Polymer glass after deformation clearly showing a reorientation of C-C bonds and Benzene rings (the obvious changes are denoted by arrows).

# Figures

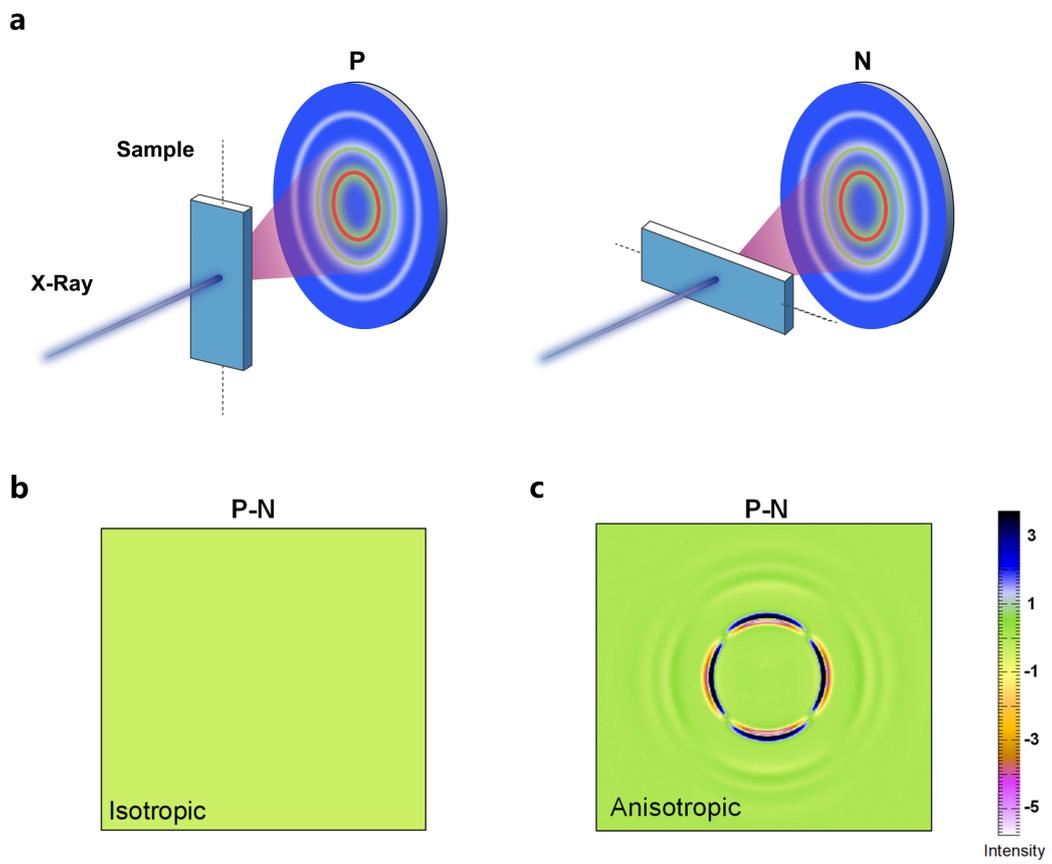

**Fig. 1 by J. Dong et al.**

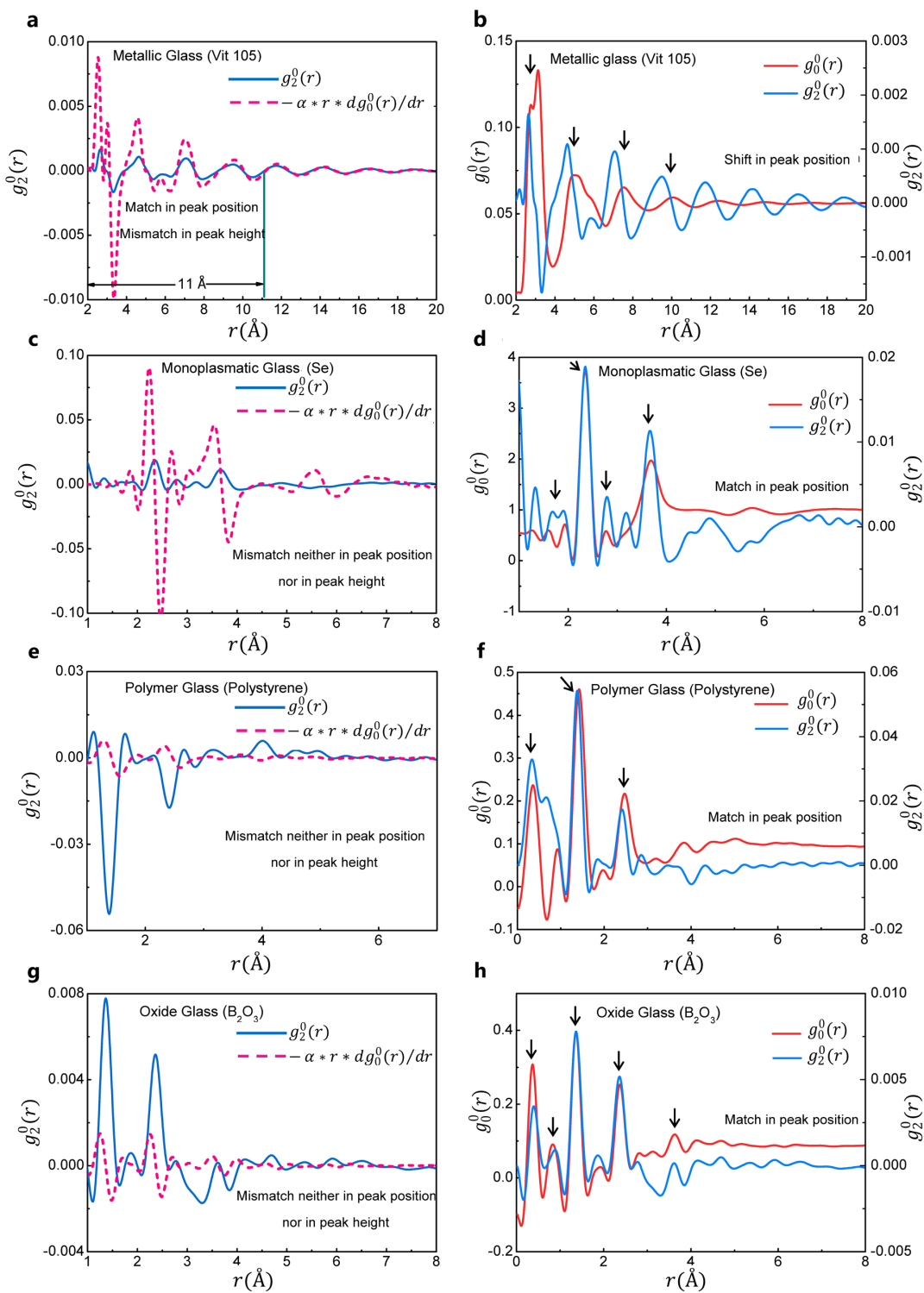

Fig. 2 by J. Dong et al.

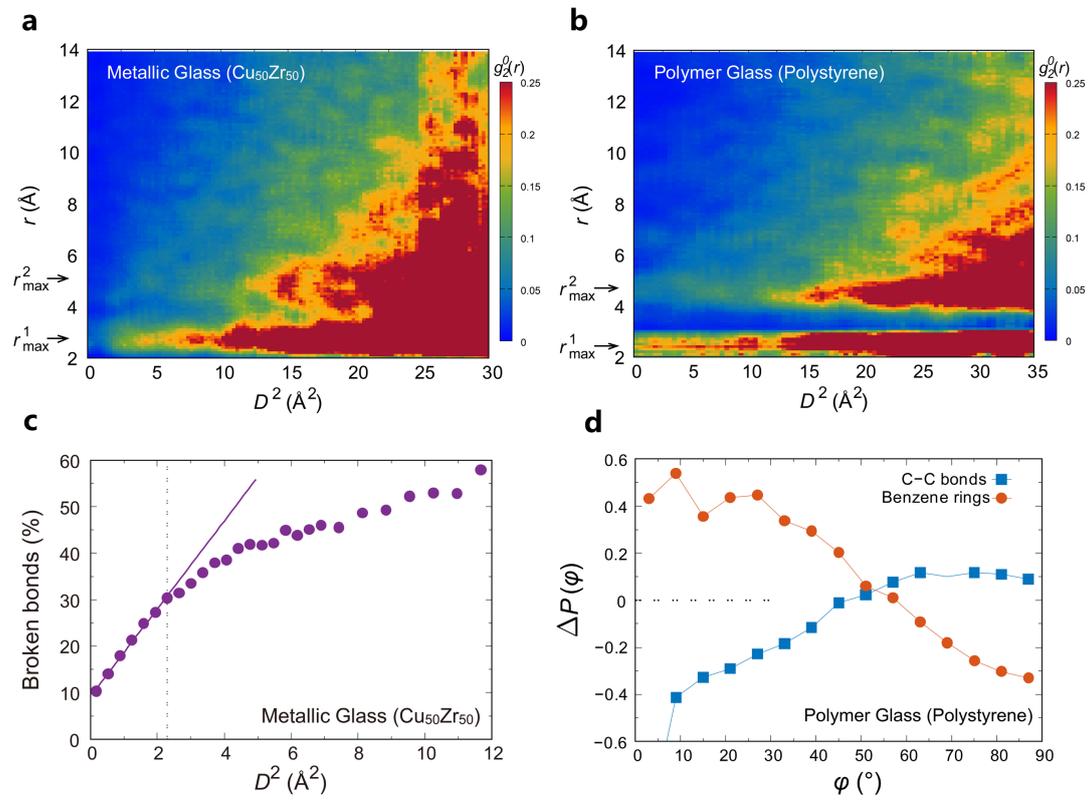

**Fig. 3 by J. Dong et al.**

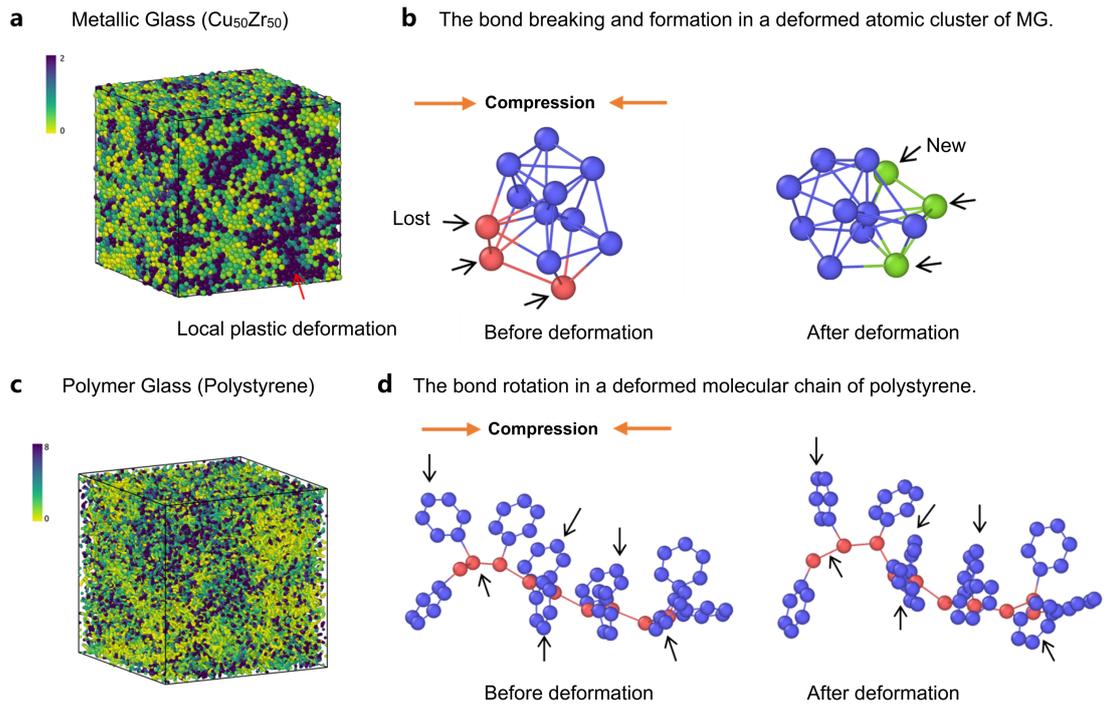

**Fig. 4** by J. Dong et al.